\documentclass[aps,pr,twocolumn,superscriptaddress,preprintnumbers,showpacs,amsmath,amssymb,floatfix]{revtex4}
\usepackage{amsmath}
\usepackage{graphicx} 
\usepackage{dcolumn}  
\usepackage{color}
\RequirePackage{xspace}
\usepackage{relsize}
\usepackage{multirow}
\usepackage{threeparttable}
\usepackage{mathtools}
\def\Y#1S{\ensuremath{\Upsilon{(#1S)}}\xspace}
\def\bbbar  {\ensuremath{b\overline b}\xspace}
\def\invfb{\ensuremath{\mbox{\,fb}^{-1}}\xspace}
\def\d{\ensuremath{\dagger}}
\def\KS  {\ensuremath{K^0_{\scriptscriptstyle S}}\xspace}
\def\antiproton{\ensuremath{\overline p}\xspace}
\def\cm   {\ensuremath{{\rm \,cm}}\xspace}
\newcommand{\gev}{\ensuremath{\mathrm{\,Ge\kern -0.1em V}}\xspace}
\newcommand{\mev}{\ensuremath{\mathrm{\,Me\kern -0.1em V}}\xspace}
\newcommand{\kev}{\ensuremath{\mathrm{\,ke\kern -0.1em V}}\xspace}
\newcommand{\gevc}{\ensuremath{{\mathrm{\,Ge\kern -0.1em V\!/}c}}\xspace}
\newcommand{\mevc}{\ensuremath{{\mathrm{\,Me\kern -0.1em V\!/}c}}\xspace}
\newcommand{\gevcc}{\ensuremath{{\mathrm{\,Ge\kern -0.1em V\!/}c^2}}\xspace}
\newcommand{\mevcc}{\ensuremath{{\mathrm{\,Me\kern -0.1em V\!/}c^2}}\xspace}
\newcommand\KorKbar{\mathrlap{\mbox{\raisebox{8pt}{\tiny(---)}}}{K}{}}

\begin{document}

\preprint{\vbox{ 
    \hbox{BELLE-CONF-1604}
    \hbox{UCHEP-16-02}
}}
\title{ \quad\\[1.0cm] Study of {\boldmath $\chi_{bJ}(1P)$} Properties in the Radiative  {\boldmath $\Y2S$} Decays}
\noaffiliation
\affiliation{Aligarh Muslim University, Aligarh 202002}
\affiliation{University of the Basque Country UPV/EHU, 48080 Bilbao}
\affiliation{Beihang University, Beijing 100191}
\affiliation{University of Bonn, 53115 Bonn}
\affiliation{Budker Institute of Nuclear Physics SB RAS, Novosibirsk 630090}
\affiliation{Faculty of Mathematics and Physics, Charles University, 121 16 Prague}
\affiliation{Chiba University, Chiba 263-8522}
\affiliation{Chonnam National University, Kwangju 660-701}
\affiliation{University of Cincinnati, Cincinnati, Ohio 45221}
\affiliation{Deutsches Elektronen--Synchrotron, 22607 Hamburg}
\affiliation{University of Florida, Gainesville, Florida 32611}
\affiliation{Department of Physics, Fu Jen Catholic University, Taipei 24205}
\affiliation{Justus-Liebig-Universit\"at Gie\ss{}en, 35392 Gie\ss{}en}
\affiliation{Gifu University, Gifu 501-1193}
\affiliation{II. Physikalisches Institut, Georg-August-Universit\"at G\"ottingen, 37073 G\"ottingen}
\affiliation{SOKENDAI (The Graduate University for Advanced Studies), Hayama 240-0193}
\affiliation{Gyeongsang National University, Chinju 660-701}
\affiliation{Hanyang University, Seoul 133-791}
\affiliation{University of Hawaii, Honolulu, Hawaii 96822}
\affiliation{High Energy Accelerator Research Organization (KEK), Tsukuba 305-0801}
\affiliation{J-PARC Branch, KEK Theory Center, High Energy Accelerator Research Organization (KEK), Tsukuba 305-0801}
\affiliation{Hiroshima Institute of Technology, Hiroshima 731-5193}
\affiliation{IKERBASQUE, Basque Foundation for Science, 48013 Bilbao}
\affiliation{University of Illinois at Urbana-Champaign, Urbana, Illinois 61801}
\affiliation{Indian Institute of Science Education and Research Mohali, SAS Nagar, 140306}
\affiliation{Indian Institute of Technology Bhubaneswar, Satya Nagar 751007}
\affiliation{Indian Institute of Technology Guwahati, Assam 781039}
\affiliation{Indian Institute of Technology Madras, Chennai 600036}
\affiliation{Indiana University, Bloomington, Indiana 47408}
\affiliation{Institute of High Energy Physics, Chinese Academy of Sciences, Beijing 100049}
\affiliation{Institute of High Energy Physics, Vienna 1050}
\affiliation{Institute for High Energy Physics, Protvino 142281}
\affiliation{Institute of Mathematical Sciences, Chennai 600113}
\affiliation{INFN - Sezione di Torino, 10125 Torino}
\affiliation{Advanced Science Research Center, Japan Atomic Energy Agency, Naka 319-1195}
\affiliation{J. Stefan Institute, 1000 Ljubljana}
\affiliation{Kanagawa University, Yokohama 221-8686}
\affiliation{Institut f\"ur Experimentelle Kernphysik, Karlsruher Institut f\"ur Technologie, 76131 Karlsruhe}
\affiliation{Kavli Institute for the Physics and Mathematics of the Universe (WPI), University of Tokyo, Kashiwa 277-8583}
\affiliation{Kennesaw State University, Kennesaw, Georgia 30144}
\affiliation{King Abdulaziz City for Science and Technology, Riyadh 11442}
\affiliation{Department of Physics, Faculty of Science, King Abdulaziz University, Jeddah 21589}
\affiliation{Korea Institute of Science and Technology Information, Daejeon 305-806}
\affiliation{Korea University, Seoul 136-713}
\affiliation{Kyoto University, Kyoto 606-8502}
\affiliation{Kyungpook National University, Daegu 702-701}
\affiliation{\'Ecole Polytechnique F\'ed\'erale de Lausanne (EPFL), Lausanne 1015}
\affiliation{P.N. Lebedev Physical Institute of the Russian Academy of Sciences, Moscow 119991}
\affiliation{Faculty of Mathematics and Physics, University of Ljubljana, 1000 Ljubljana}
\affiliation{Ludwig Maximilians University, 80539 Munich}
\affiliation{Luther College, Decorah, Iowa 52101}
\affiliation{University of Maribor, 2000 Maribor}
\affiliation{Max-Planck-Institut f\"ur Physik, 80805 M\"unchen}
\affiliation{School of Physics, University of Melbourne, Victoria 3010}
\affiliation{Middle East Technical University, 06531 Ankara}
\affiliation{University of Miyazaki, Miyazaki 889-2192}
\affiliation{Moscow Physical Engineering Institute, Moscow 115409}
\affiliation{Moscow Institute of Physics and Technology, Moscow Region 141700}
\affiliation{Graduate School of Science, Nagoya University, Nagoya 464-8602}
\affiliation{Kobayashi-Maskawa Institute, Nagoya University, Nagoya 464-8602}
\affiliation{Nara University of Education, Nara 630-8528}
\affiliation{Nara Women's University, Nara 630-8506}
\affiliation{National Central University, Chung-li 32054}
\affiliation{National United University, Miao Li 36003}
\affiliation{Department of Physics, National Taiwan University, Taipei 10617}
\affiliation{H. Niewodniczanski Institute of Nuclear Physics, Krakow 31-342}
\affiliation{Nippon Dental University, Niigata 951-8580}
\affiliation{Niigata University, Niigata 950-2181}
\affiliation{University of Nova Gorica, 5000 Nova Gorica}
\affiliation{Novosibirsk State University, Novosibirsk 630090}
\affiliation{Osaka City University, Osaka 558-8585}
\affiliation{Osaka University, Osaka 565-0871}
\affiliation{Pacific Northwest National Laboratory, Richland, Washington 99352}
\affiliation{Panjab University, Chandigarh 160014}
\affiliation{Peking University, Beijing 100871}
\affiliation{University of Pittsburgh, Pittsburgh, Pennsylvania 15260}
\affiliation{Punjab Agricultural University, Ludhiana 141004}
\affiliation{Research Center for Electron Photon Science, Tohoku University, Sendai 980-8578}
\affiliation{Research Center for Nuclear Physics, Osaka University, Osaka 567-0047}
\affiliation{Theoretical Research Division, Nishina Center, RIKEN, Saitama 351-0198}
\affiliation{RIKEN BNL Research Center, Upton, New York 11973}
\affiliation{Saga University, Saga 840-8502}
\affiliation{University of Science and Technology of China, Hefei 230026}
\affiliation{Seoul National University, Seoul 151-742}
\affiliation{Shinshu University, Nagano 390-8621}
\affiliation{Showa Pharmaceutical University, Tokyo 194-8543}
\affiliation{Soongsil University, Seoul 156-743}
\affiliation{University of South Carolina, Columbia, South Carolina 29208}
\affiliation{Stefan Meyer Institute for Subatomic Physics, Vienna 1090}
\affiliation{Sungkyunkwan University, Suwon 440-746}
\affiliation{School of Physics, University of Sydney, New South Wales 2006}
\affiliation{Department of Physics, Faculty of Science, University of Tabuk, Tabuk 71451}
\affiliation{Tata Institute of Fundamental Research, Mumbai 400005}
\affiliation{Excellence Cluster Universe, Technische Universit\"at M\"unchen, 85748 Garching}
\affiliation{Department of Physics, Technische Universit\"at M\"unchen, 85748 Garching}
\affiliation{Toho University, Funabashi 274-8510}
\affiliation{Tohoku Gakuin University, Tagajo 985-8537}
\affiliation{Department of Physics, Tohoku University, Sendai 980-8578}
\affiliation{Earthquake Research Institute, University of Tokyo, Tokyo 113-0032}
\affiliation{Department of Physics, University of Tokyo, Tokyo 113-0033}
\affiliation{Tokyo Institute of Technology, Tokyo 152-8550}
\affiliation{Tokyo Metropolitan University, Tokyo 192-0397}
\affiliation{Tokyo University of Agriculture and Technology, Tokyo 184-8588}
\affiliation{University of Torino, 10124 Torino}
\affiliation{Toyama National College of Maritime Technology, Toyama 933-0293}
\affiliation{Utkal University, Bhubaneswar 751004}
\affiliation{Virginia Polytechnic Institute and State University, Blacksburg, Virginia 24061}
\affiliation{Wayne State University, Detroit, Michigan 48202}
\affiliation{Yamagata University, Yamagata 990-8560}
\affiliation{Yonsei University, Seoul 120-749}
  \author{A.~Abdesselam}\affiliation{Department of Physics, Faculty of Science, University of Tabuk, Tabuk 71451} 
  \author{I.~Adachi}\affiliation{High Energy Accelerator Research Organization (KEK), Tsukuba 305-0801}\affiliation{SOKENDAI (The Graduate University for Advanced Studies), Hayama 240-0193} 
  \author{K.~Adamczyk}\affiliation{H. Niewodniczanski Institute of Nuclear Physics, Krakow 31-342} 
  \author{H.~Aihara}\affiliation{Department of Physics, University of Tokyo, Tokyo 113-0033} 
  \author{S.~Al~Said}\affiliation{Department of Physics, Faculty of Science, University of Tabuk, Tabuk 71451}\affiliation{Department of Physics, Faculty of Science, King Abdulaziz University, Jeddah 21589} 
  \author{K.~Arinstein}\affiliation{Budker Institute of Nuclear Physics SB RAS, Novosibirsk 630090}\affiliation{Novosibirsk State University, Novosibirsk 630090} 
  \author{Y.~Arita}\affiliation{Graduate School of Science, Nagoya University, Nagoya 464-8602} 
  \author{D.~M.~Asner}\affiliation{Pacific Northwest National Laboratory, Richland, Washington 99352} 
  \author{T.~Aso}\affiliation{Toyama National College of Maritime Technology, Toyama 933-0293} 
  \author{H.~Atmacan}\affiliation{Middle East Technical University, 06531 Ankara} 
% ...
\collaboration{The Belle Collaboration}
\begin{abstract}
  We report a study of radiative decays of $\chi_{bJ}(1P)(J=0,1,2)$ mesons into 
  74 hadronic final states comprising charged and neutral pions, kaons, protons;
  out of these, 41 modes are observed with at least 5 standard deviation 
  significance. Our measurements not only improve the previous measurements by 
  the CLEO Collaboration but also lead to first observations in many new modes.
  The large sample allows us to probe the total decay width of the 
  $\chi_{b0}(1P)$. 
  In the absence of a statistically significant result, a $90\%$ 
  confidence-level upper limit is set on the width at 
  $\Gamma_{\rm total}< 2.4\mev$. 
  Our results are based on $24.7\invfb$ of $e^{+}e^{-}$ collision data recorded 
  by the Belle detector at the $\Y2S$ resonance, corresponding to 
  $(157.8\pm3.6)\times10^6$ $\Y2S$ decays.
\end{abstract}


\maketitle
\section{Introduction}
\label{sec:intro}
The $P$-wave spin triplet states of bottomonia, $\chi_{bJ}(nP)$, are copiously 
produced from the $\Upsilon(nS)$ states through electric dipole radiative 
transitions. 
The $\chi_{b0}(1P)$ and $\chi_{b2}(1P)$ states, with positive parity and 
charge-parity, can annihilate to two real photons or gluons. 
Various theoretical predictions for the two-gluon widths 
($\Gamma_{\rm total}\approx\Gamma_{2g}$) of the 
$\chi_{b0}(1P)$ ($J^{\rm PC}= 0^{++}$) and $\chi_{b2}(1P)$ ($J^{\rm PC}= 2^{++}$) 
states are listed in Table~\ref{tab:width}. 
All calculations predict the width of $\chi_{b0}(1P)$ to be larger than that of
$\chi_{b2}(1P)$, reaching above $2~\mev$. 
On the experimental front, so far there is no measurement for the width of any 
of the $\chi_{bJ}(1P)$ states. 
According to the Landau-Yang theorem~\cite{LYT}, a $J$ = 1 particle cannot decay
to two identical massless spin-1 particles, thus the $\chi_{b1}(1P)$ cannot 
decay to two real gluons or photons. 
However, this process is allowed if one of the gluons is virtual giving rise to
a quark-antiquark pair~\cite{HeavyQuark}.  

In 2008, the CLEO Collaboration reported the first observations of 
$\chi_{bJ}(1P)$ and $\chi_{bJ}(2P)$ decays into specific final states of light 
hadrons, where these $P$-wave states are produced in radiative transitions of 
the \Y2S and \Y3S resonances, respectively~\cite{chibj1p:cleo}. 
In our earlier search for a new state near 9975~\mevcc, 
$X_{\bbbar}(9975)$~\cite{xbb:belle}, we observed a large signal yield for 
$\chi_{bJ}(1P)$ states from the sum of 26 exclusive hadronic final states. 
The yields for $\chi_{bJ}(1P)(J=0,1,2)$ were $299 \pm 22$, $946 \pm 36$ and 
$582 \pm 31$, respectively. This motivated a study of the product branching 
fractions 
${\cal B}[\Y2S\to\gamma\chi_{bJ}(1P)]\times{\cal B}[\chi_{bJ}(1P)\to h_{i}]$, 
where $h_{i}$ is a specific hadronic mode; such decays of $\chi_{bJ}(1P)$ mesons
give us insight into how initial quarks and gluons hadronize~\cite{patrignani1}.
Our study, using the world's largest $e^{+}e^{-}\to$\Y2S data sample, permits us
not only to improve upon the earlier measurements of $14$ 
modes~\cite{chibj1p:cleo}, but also to uncover many new modes. 
The analysis also provides an opportunity for a width measurement of 
$\chi_{b0}(1P)$, which is predicted to be the largest among the three 
$\chi_{bJ}(1P)$ states.

\begin{table}[htb]
  \caption{Predicted total widths (in $\kev$) of the $\chi_{b0}(1P)$ and 
    $\chi_{b2}(1P)$ states, assuming $\Gamma_{\rm total}\approx\Gamma_{2g}$.}
  \label{tab:width}
  \begin{threeparttable}
    \begin{tabular}
      {@{\hspace{0.35cm}}l@{\hspace{0.35cm}}  @{\hspace{0.35cm}}c@{\hspace{0.35cm}}
        @{\hspace{0.35cm}}l@{\hspace{0.35cm}} }
      \hline \hline
      $\Gamma [\chi_{b0}(1P)]$ ($\kev$) & $\Gamma [\chi_{b2}(1P)]$ ($\kev$)& Ref.\\
      \hline
      $2.03$ ($\mev$) & $122.84$ & \cite{width:segovia} \\
      $431^{+45}_{-49}$ & $214^{+1}_{-0}$ & \cite{width:hwang} \\
      $887$ & $220$ &  \cite{width:wang1,width:wang2}\\
      $960\,(2740)$ & $330\,(250)$ & \cite{width:laverty, width:hwang}\\
      $653$ & $109$ & \cite{width:ebert}\\
      $2150\,(2290)$ & $220\,(330)$ & \cite{width:gupta}\\
      $672$ & $123$ & \cite{width:godfrey}\\
      \hline \hline
    \end{tabular}
    \begin{tablenotes}
      \footnotesize{
      \item The values of ~\cite{width:laverty} are obtained by the perturbative (nonperturbative) calculation and that for ~\cite{width:gupta} are obtained by the QCD potential (alternative treatment)}
    \end{tablenotes}
  \end{threeparttable}
\end{table}

\section{Data and Simulation Samples}
\label{sec:data}
We perform this study using a $24.7\invfb$ data sample, equivalent to 
$(157.8\pm3.6)\times10^{6}$ $\Y2S$ events~\cite{NY2S}, collected at the 
$\Y2S$ resonance with the Belle detector~\cite{belle:detector} at the KEKB 
asymmetric-energy $e^+e^-$ collider~\cite{KEKB}.
A $1.7\invfb$ data sample, recorded $30\mev$ below the $\Y2S$ peak, provides a 
control sample to study the $e^+e^-\to q\overline{q}$ $(q=u,d,s,c)$ continuum
background.

Half a million signal Monte Carlo (MC) events are produced for each final state
studied.  
The radiative transitions from the $\Y2S$ are generated using the helicity 
amplitude formalism~\cite{mc:helamp}. 
Hadronic decays of $\chi_{bJ}$ are modeled assuming a phase space distribution, 
where an interface to PHOTOS~\cite{mc:photos} has been added to incorporate 
final state radiation effects. 
As all signal MC samples are generated with a phase space distribution, 
possible intermediate decays such as  
$\rho^{0}\to\pi^{+}\pi^{-}$, $\rho^{\pm}\to\pi^{\pm}\pi^{0}$, 
$\phi\to K^{+}K^{-}$, $\omega\to\pi^{+}\pi^{-}\pi^{0}$, 
$\KorKbar^*(892)^0 \to K^{\pm}\pi^{\mp}/\KS\pi^{0}$and 
$K^{\star}(892)^{\pm}\to\KS\pi^{\pm}/ K^{\pm}\pi^{0}$ 
are considered primarily when we later estimate systematic uncertainties in 
efficiency.
Inclusive $\Y2S$ MC events, produced using \textsc{Pythia}~\cite{mc:pythia} 
with the same luminosity as the data, are utilized for background studies.

\section{Experimental Apparatus}
\label{sec:belle}
The Belle detector~\cite{belle:detector} is a large-solid-angle magnetic
spectrometer consisting of a silicon vertex detector,
a 50-layer central drift chamber (CDC), an array of
aerogel threshold Cherenkov counters (ACC), 
a barrel-like arrangement of time-of-flight
scintillation counters (TOF), and an electromagnetic calorimeter
comprised of CsI(Tl) crystals (ECL). All these detector elements are
located inside 
a superconducting solenoid coil that provides a 1.5~T
magnetic field.  An iron flux-return located outside
the coil is instrumented with resistive plate chambers 
to detect $K_L^0$ mesons and muons.  

\section{{\boldmath $\Y2S$} Reconstruction}
\label{sec:Y2S}
Reconstruction begins with the selection of $\pi^{\pm}$, $K^{\pm}$, 
$p/\antiproton$, $\KS$ and $\pi^{0}$ to reconstruct a $\chi_{bJ}(1P)$ candidate.
We then select a $\gamma$ candidate and combine it with the $\chi_{bJ}(1P)$ to 
form a $\Y2S$ candidate.

\subsection{{\boldmath $\chi_{bJ}(1P)$} Reconstruction}
\label{sec:Y2S:chibj}
The $\chi_{bJ}(1P)$ state can decay to many hadronic final states. For decays into all-charged final states, we focus on the same $26$ modes as in Ref.~\cite{xbb:belle}: \\
\begin{minipage}{0.48\textwidth}  
  $2(\pi^+\pi^-)$, $3(\pi^+\pi^-)$, $4(\pi^+\pi^-)$, $5(\pi^+\pi^-)$, 
  $\pi^+\pi^-K^+K^-$, $2(\pi^+\pi^-)K^+K^-$, $3(\pi^+\pi^-)K^+K^-$, 
  $4(\pi^+\pi^-)K^+K^-$, $2(K^+K^-)$, $\pi^+\pi^-2(K^+K^-)$, 
  $2(\pi^+\pi^-K^+K^-)$, $3(\pi^+\pi^-)2(K^+K^-)$, $\pi^+\pi^-p\antiproton$, 
  $2(\pi^+\pi^-)p\antiproton$, $3(\pi^+\pi^-)p\antiproton$, 
  $4(\pi^+\pi^-)p\antiproton$, $\pi^+\pi^-K^+K^-p\antiproton$, 
  $2(\pi^+\pi^-)K^+K^-p\antiproton$, $3(\pi^+\pi^-)K^+K^-p\antiproton$, 
  $\pi^{\pm}K^{\mp}\KS$, $\pi^+\pi^-\pi^{\pm}K^{\mp}\KS$, 
  $2(\pi^+\pi^-)\pi^{\pm}K^{\mp}\KS$, $3(\pi^+\pi^-)\pi^{\pm}K^{\mp}\KS$, 
  $\pi^+\pi^-2\KS$, $2(\pi^+\pi^-\KS)$, and $3(\pi^+\pi^-)2\KS$.
\end{minipage}
One $\pi^{0}$ is added to the above final states, excluding 
$2(\pi^+\pi^-)\pi^0$, $3(\pi^+\pi^-)\pi^0$, $4(\pi^+\pi^-)\pi^0$ and 
$5(\pi^+\pi^-)\pi^0$ as those are forbidden by G-parity 
conservation~\cite{chibj1p:cleo}, resulting in a total of 22 modes 
reconstructed with one $\pi^0$. 
An additional 26 modes are reconstructed with the addition of two $\pi^{0}$'s 
to the charged final states enumerated above.

In total, 74 light hadronic decay modes of the $\chi_{bJ}(1P)$ are 
reconstructed. Charged particles ($\pi^{\pm}$, $K^{\pm}$, $p/\antiproton$), 
$\KS$ and $\pi^0$ mesons are selected as follows:

{\bf Impact parameters:} For charged tracks, we require maximum distances of 
closest approach with respect to the interaction point (IP), in both the $xy$ 
transverse plane (`$dr$') and along the $z$ axis (`$|dz|$'), with the $z$-axis 
defined as the direction opposite to the $e^+$ beam. The selection criteria 
applied are  $dr<1\cm$ and $|dz|<4\cm$, to ensure that charged tracks are 
originating from the IP and not the result of beam-wall or beam-gas 
interactions. These requirements are not imposed on the charged pions arising 
from a $\KS$ decay.

{\bf Number of charged tracks:} Selection criteria are also applied to the 
number of tracks ($4$, $6$, $8$, $10$, or $12$) depending on the mode.

{\bf Charged pion and kaon selection:} Charged pions and kaons are identified 
based on their likelihood ratios 
${\cal L}_{K/\pi}=\frac{{\cal L}_{K}}{{\cal L}_{K}+{\cal L}_{\pi}}$,  where 
${\cal L}_{K}$ and ${\cal L}_{\pi}$ are the likelihoods for $K^{\pm}$ and 
$\pi^{\pm}$, respectively, calculated using the number of photoelectrons from 
the ACC, information from the TOF and specific ionization in the CDC.
We apply ${\cal L}_{K/\pi}$ $<0.6$ for selecting pions and also require them not
to be a daughter of any $\KS$ candidate. Similarly, ${\cal L}_{K/\pi}$ $>0.6$ is
used to select kaons. For the above mentioned criteria, the kaon identification
efficiency is 81 -- 90\% with a kaon-to-pion misidentification probability of 
9 -- 14\%. Pions are detected with an efficiency of 91 -- 95\% with a 
kaon-to-pion misidentification probability of 8 -- 13\%.

{\bf Selection of {\boldmath $p/\antiproton$}:} protons and antiprotons are 
identified based on the likelihood ratios ${\cal L}_{p/K}$ and ${\cal L}_{p/\pi}$.
The criteria applied for proton or antiproton selection are: 
${\cal L}_{p/K}$ $>0.7$ and ${\cal L}_{p/\pi}$  $>0.7$. The proton identification
efficiency is 95\%, while the probability of a kaon being misidentified as a 
proton is below 3\%.

{\bf  {\boldmath $\KS$} selection:}  Candidate $\KS$ mesons are reconstructed by
combining two oppositely charged tracks (pion mass assumed) with an invariant 
mass between $486$ and $509$ $\mevcc$. The selected candidates are also required
to satisfy the criteria described in Ref.~\cite{ks_selection} to ensure that 
their decay vertices are displaced from the IP.

{\bf {\boldmath $\pi^0$} reconstruction:} A $\pi^0$ is reconstructed from a pair
of $\gamma$'s, each having energy greater than $100\mev$. The reconstructed 
$\pi^0$ should have an invariant mass within $[113,157]\mevcc$, which is 
$\pm 3.5 \sigma$ around the nominal $\pi^0$ mass~\cite{PDG}. 

\subsection{Transition Photon Selection}
\label{sec:Y2S:gamma}

Radiative photon candidates arising from the transition 
$\Y2S\to\gamma\chi_{bJ}(1P)$ are chosen based on the following quantities:

{\bf Energy of the photon {\boldmath ($E_{\gamma}$)}:} The signal photon has 
energy 100 -- 240\mev for $\chi_{b0}(1P)$, 70 -- 190\mev for $\chi_{b1}(1P)$, 
and 50 -- 170\mev for $\chi_{b2}(1P)$. 
To cover all three ranges, a lower threshold of 30\mev is applied on $E_{\gamma}$
for selecting the photon.

{\bf Polar angle of the {\boldmath $\gamma$ ($\theta_{\gamma}$)}:} We define 
$\theta_\gamma$ as the angle between the $\gamma$ direction and the $z$ axis. 
To reduce the beam background contamination, photons only in the barrel region 
($\theta_{\gamma}~\in~[0.5,~2.3]\rm~rad$) are retained.

{\bf E9/E25:} The showers in the ECL have a variety of energy deposition 
patterns; the variable E9/E25 compares the amount of energy deposited in a 
$3\times3$ crystal block (E9) around the crystal with maximum energy to that in
a $5\times5$ crystal block (E25). 
We apply the criterion E9/E25 $> 0.85$ to select photon candidates.

{\bf Charged track match:} The photon cluster in the ECL should not match with 
the trajectories of charged track(s) in the CDC extrapolated into the ECL. 

{\bf {\boldmath $\pi^{0}$} rejection:} The candidate $\gamma$ should be 
inconsistent with those arising from the decay of $\pi^{0}$ in the final state.

Selected $\gamma$ candidates satisfying the above criteria are then combined 
with the reconstructed $\chi_{bJ}(1P)$ candidate to form an $\Y2S$ candidate. 
At this stage, we apply a very loose requirement on $|\Delta E|<0.5\gev$, 
where $\Delta E$ is the difference between the energy of the $\Y2S$ candidate 
and the center-of-mass energy.

\subsection{Continuum Suppression}
\label{sec:Y2S:cont}
Continuum background events result from light ($u$, $d$, $s$ and $c$) 
quark-antiquark pairs produced in $e^+e^-$ collisions. 
Signal $\chi_{bJ}(1P)$ decay events have a spherical topology, 
in contrast to `back-to-back' jetlike continuum events. 
To suppress the latter, the cosine of the angle between the photon candidate 
and the thrust axis (calculated from the final state hadrons) in the $\Y2S$ 
rest frame, $\cos\theta_{T}$, is considered.
Signal events have a uniform distribution in this variable while continuum 
events peak near $|\cos\theta_{T}|=1$. 
A requirement  $|\cos\theta_{T}|<0.8$ is therefore applied to reduce 
the continuum background.

\section{\bf Kinematic fit and Reduced {\boldmath $\chi^2$} criterion}
\label{sec:KinFit}
In an analysis where all the final state particles are reconstructed, a 
kinematic fit applying energy-momentum conservation (4C) can be helpful in 
improving the mass resolution. The reduced $\chi^2$ ($\chi^2/{\rm NDF}$) from 
the 4C fit is also used as a selection criterion for signal (and to select the 
best candidate in cases where multiple candidates are found). After the event 
selection criteria and continuum suppression described above are imposed, a
criterion on the reduced $\chi^2$ is applied for further background suppression.

Optimization for the criterion on the reduced $\chi^2$ is performed for the 
$\chi_{b0}(1P)$, as it is the signal of interest for the width measurement. 
The optimization is performed using either (a) sum of the product 
branching-fractions 
${\cal B}[\Y2S\to\gamma\chi_{b1}(1P)]\times{\cal B}[\chi_{b1}(1P)\to X_{i}]$ for 
$13$ modes (available in Ref~\cite{chibj1p:cleo}, excluding upper limits), 
multiplied by the ratio of the sum of signal yields for $\chi_{b0}(1P)$ and 
$\chi_{b1}(1P)$ in those $13$ modes, or (b) the product branching fraction as 
in case (a), but varied by $\pm1\sigma$ around the obtained uncertainty.
The $\chi_{b0}(1P)$ signal region in terms of 
$\Delta M\equiv M[\chi_{bJ}(1P)\gamma]-M[\chi_{bJ}(1P)]$ distribution is defined
as $[138,180]\mevcc$, which corresponds to a $\pm3\sigma$ window. The 
non-continuum background contribution is estimated from a data-equivalent MC 
sample of $\Y2S$ decaying generically, in the same mass window. 

\begin{figure}[htb]
  \includegraphics[width=0.45\textwidth]{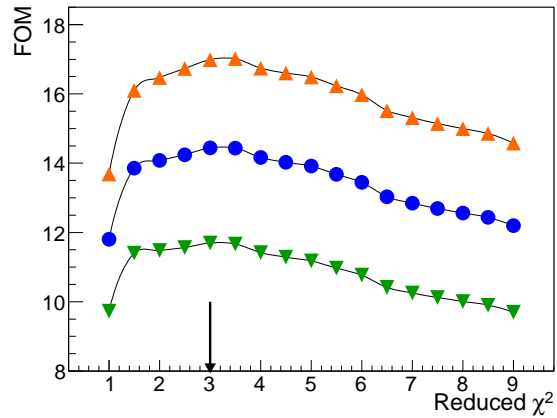}
  \caption{Variation of the FOM with the reduced $\chi^2$ cut. The FOM is also 
    recalculated by varying the product branching fraction by $\pm 1\sigma$ 
    around its error. The curve with upward orange triangles is for the 
    positive variation, the curve with downward green triangles for the 
    negative variation, whereas the curve with blue circles corresponds to 
    the central value.}
  \label{fig:fom}
\end{figure}

A figure-of-merit (FOM) is calculated as $S/\sqrt{(S+B)}$, where $S$ ($B$) is 
the expected signal (background) yield. The dependence of the FOM on the 
reduced $\chi^2$ condition is shown in Fig.~\ref{fig:fom}, from which we select
a criteria on reduced $\chi^2 < 3$. The FOM is recalculated for the case (b) 
scenario, and is again shown in Fig.~\ref{fig:fom}. The optimal point remains 
unchanged.

\section{Fit to data}
\label{sec:modes}
We use MC simulation to set up an extended unbinned maximum-likelihood fit of 
the three $\chi_{bJ}(1P)(J=0,1,2)$ candidate $\Delta M$ distributions. 
To decide the probability distribution function (PDF) for the signal components,
the respective signal MC samples are studied. 
All three signal shapes are parametrized with the sum of a symmetric and an 
asymmetric Gaussian function having common mean. The asymmetric component is 
added to account for low-energy tails. 
The (three) width parameters of the Gaussians are fixed to the MC values for 
all modes in the fit. 
The width of the symmetric Gaussians for $\chi_{bJ}(1P)(J=0,1,2)$ are found to be $4.6 \pm 0.5 \mev$, $4.2 \pm 0.4 \mev$ and $4.1 \pm 0.4 \mev$, respectively. 
The parameters for all modes in MC are close and have a small RMS. 
Therefore, we fix each parameter to its mean value and vary within the RMS to 
estimate the assorted systematic uncertainty.
In order to account for a modest difference in the detector resolution 
between data and simulations, we apply a calibration factor common to the 
three signal components. 
The background is modeled with the sum of an exponential function and a 
first-order Chebyshev polynomial. 
The corresponding four parameters (exponent, slope, relative fraction and 
background yield) are allowed to vary in the fit.

\begin{figure}[htb]
  \includegraphics[width=0.5\textwidth]{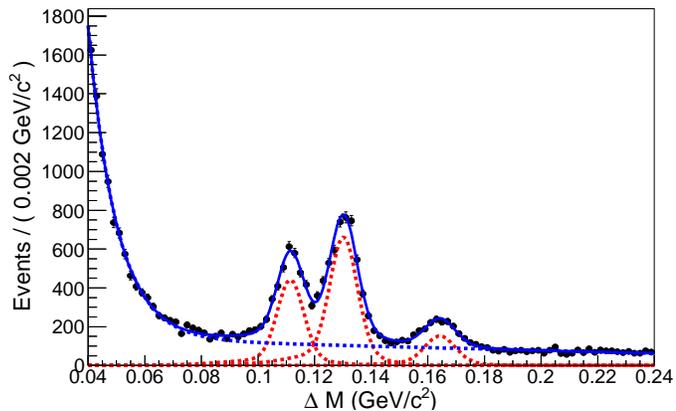}
  \caption{$\Delta M$ distribution in the $\Y2S$ data. Black points with error bars are the data; the solid blue and dashed blue curves show the total fit and background components. The three $\chi_{bJ}(1P)$ ($J=0,1,$ and 2, respectively from right to left) components are indicated by the red dashed curves.}
  \label{fig:fitdt}
\end{figure}

The result of the likelihood fit to the $\Delta M$ distributions for the sum of
$74$ modes in data is shown in Fig.~\ref{fig:fitdt}; the $p$-value of the fit 
is 0.37. 
The $\chi_{bJ}(1P)$ signal yields are found to be 5 times larger than 
those obtained in our previous analysis~\cite{xbb:belle}. The masses of the 
$\chi_{bJ}(1P)(J=0,1,2)$ states obtained are in excellent agreement with their 
world average values~\cite{PDG}. Results are summarized in Table~\ref{tab:cj}, 
including the obtained signal yields in each mode. The resolution calibration 
factor is found to be $1.13\pm0.02$, in a reasonable agreement with previous 
Belle estimates for this correction~\cite{xbb:belle}.

\begin{table}[htb]
  \caption{$\chi_{bJ}(1P)(J=0,1,2)$ masses and yields obtained from the fit to 
    data. (Errors are statistical only.)}
  \label{tab:cj}
  \begin{tabular}
    {@{\hspace{0.35cm}}l@{\hspace{0.35cm}}  @{\hspace{0.35cm}}c@{\hspace{0.35cm}}
      @{\hspace{0.35cm}}l@{\hspace{0.35cm}}}
\hline \hline
Signal & Yield & Mass in $\mevcc$\\
\hline
$\chi_{b0}(1P)$ & $1197\pm 58$ & $9858.98\pm0.33$\\
$\chi_{b1}(1P)$ & $4747\pm 94$ & $9893.05\pm0.12$ \\
$\chi_{b2}(1P)$ & $3064\pm 87$ & $9911.81\pm0.16$\\
\hline \hline
\end{tabular}
\end{table}

\subsection{Mode Selection}
\label{sec:modes:modeSel}

\begin{figure*}[htp]
  \includegraphics[width=0.9\textwidth]{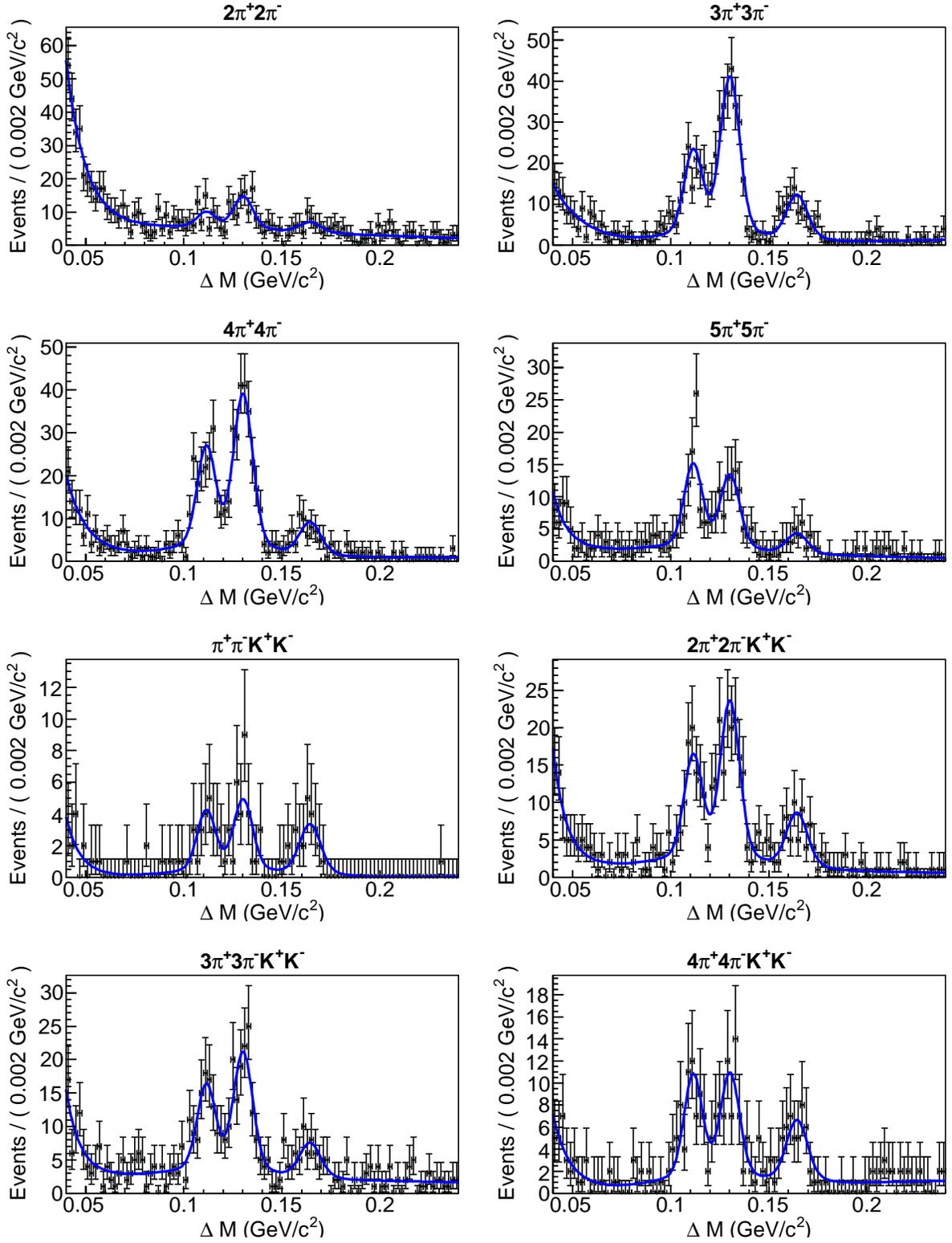}
  \caption{$\Delta M$ distributions in $\Y2S$ data for $2(\pi^+\pi^-)$, $3(\pi^+\pi^-)$, $4(\pi^+\pi^-)$, $5(\pi^+\pi^-)$, $\pi^+\pi^-K^+K^-$, $2(\pi^+\pi^-)K^+K^-$, $3(\pi^+\pi^-)K^+K^-$, and $4(\pi^+\pi^-)K^+K^-$ final states. Black dots with error bars are data and the blue curves represent the total fit result.}
  \label{fig:modes1}
\end{figure*}

\begin{figure}[htb]
  \includegraphics[width=0.9\textwidth]{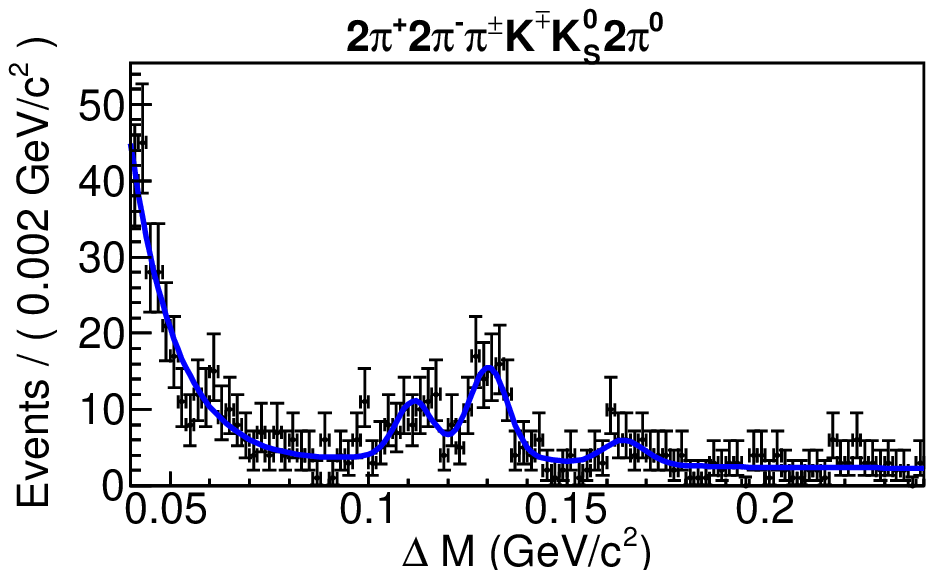}
  \caption{$\Delta M$ distributions in $\Y2S$ data for $2(\pi^+\pi^-)\pi^{\pm}K^{\mp}\KS 2\pi^0$ final state. Black dots with error bars are data and the blue curve represents the total fit result.}
  \label{fig:modes6}
\end{figure}

With signal shapes, including means, fixed in the fit to the result obtained 
from the sum of 74 modes (described previously), we fit the $\Delta M$ 
distribution in each individual mode to determine its significance as 
$\sqrt{-2\ln({\cal L}_{0}/{\cal L}_{\rm max})}$, where 
${\cal L}_{0}$ (${\cal L}_{\rm max}$) is the likelihood value when the signal 
yield is fixed to zero (allowed to vary). In total, 41 modes have above $5$ 
standard deviation ($\sigma$) significance in at least one of the 
$\chi_{bJ}(1P) (J=0,1,2)$ signals. 
The $\Delta M$ distribution in those 41 modes is shown in 
Figs.\,\ref{fig:modes1} --\,\ref{fig:modes6}.

\begin{figure*}[htp]
  \includegraphics[width=0.9\textwidth]{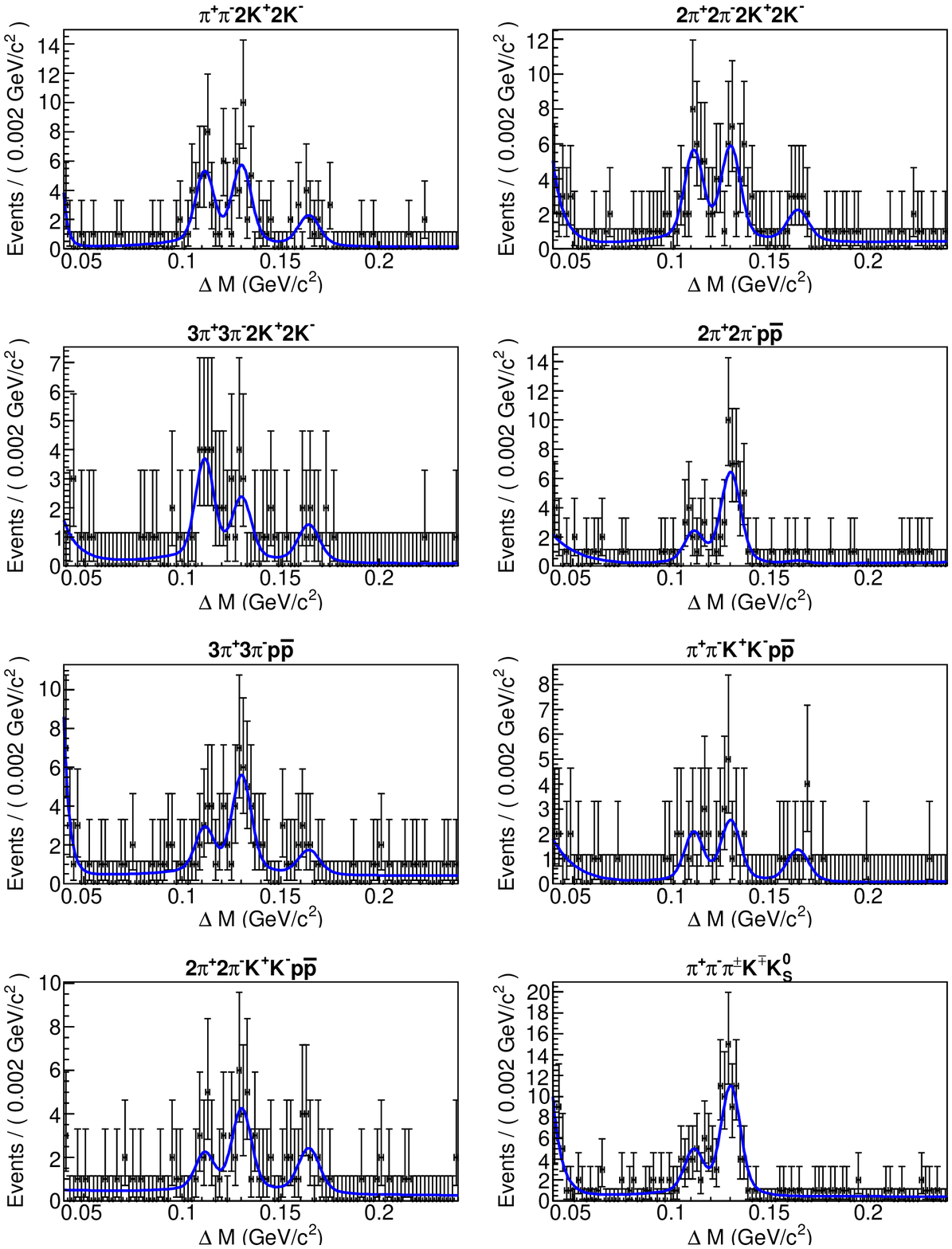}
  \caption{$\Delta M$ distributions in $\Y2S$ data for $\pi^+\pi^-2(K^+K^-)$, $2(\pi^+\pi^-K^+K^-)$, $3(\pi^+\pi^-)2(K^+K^-)$, $2(\pi^+\pi^-)p\antiproton$, $3(\pi^+\pi^-)p\antiproton$, $\pi^+\pi^-K^+K^-p\antiproton$, $2(\pi^+\pi^-)K^+K^-p\antiproton$, and $\pi^+\pi^-\pi^{\pm}K^{\mp}\KS$ final states. Black dots with error bars are data and the blue curves represent the total fit result.}
  \label{fig:modes2}
\end{figure*}

\begin{figure*}[htp]
  \includegraphics[width=0.9\textwidth]{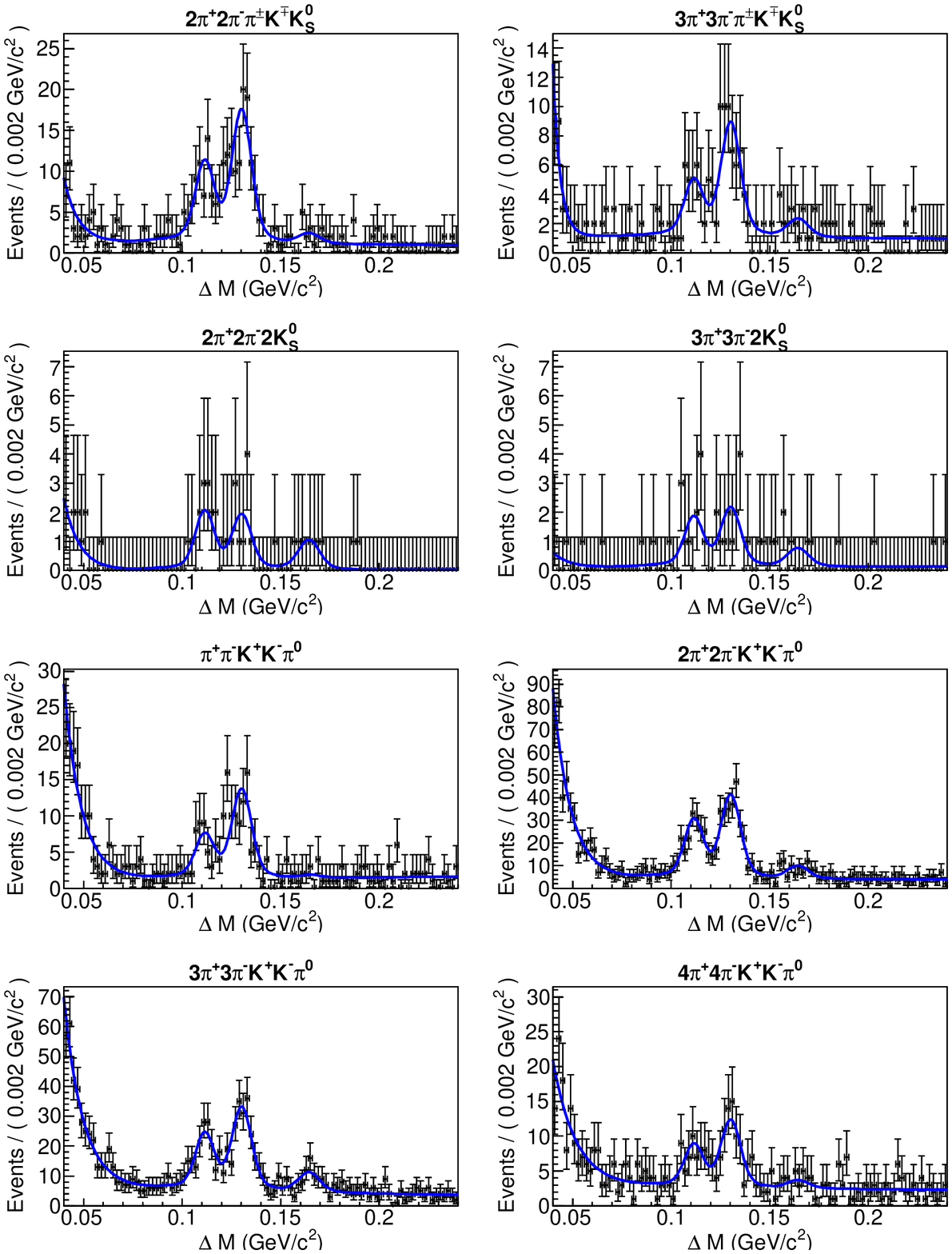}
  \caption{$\Delta M$ distributions in $\Y2S$ data for  $2(\pi^+\pi^-)\pi^{\pm}K^{\mp}\KS$, $3(\pi^+\pi^-)\pi^{\pm}K^{\mp}\KS$, $2(\pi^+\pi^-\KS)$, $3(\pi^+\pi^-)2\KS$, $\pi^+\pi^-K^+K^-\pi^0$, $2(\pi^+\pi^-)K^+K^-\pi^0$, $3(\pi^+\pi^-)K^+K^-\pi^0$ and $4(\pi^+\pi^-)K^+K^-\pi^0$ final states. Black dots with error bars are data and the blue curves represent the total fit result.}
  \label{fig:modes3}
\end{figure*}

\begin{figure*}[htp]
  \includegraphics[width=0.9\textwidth]{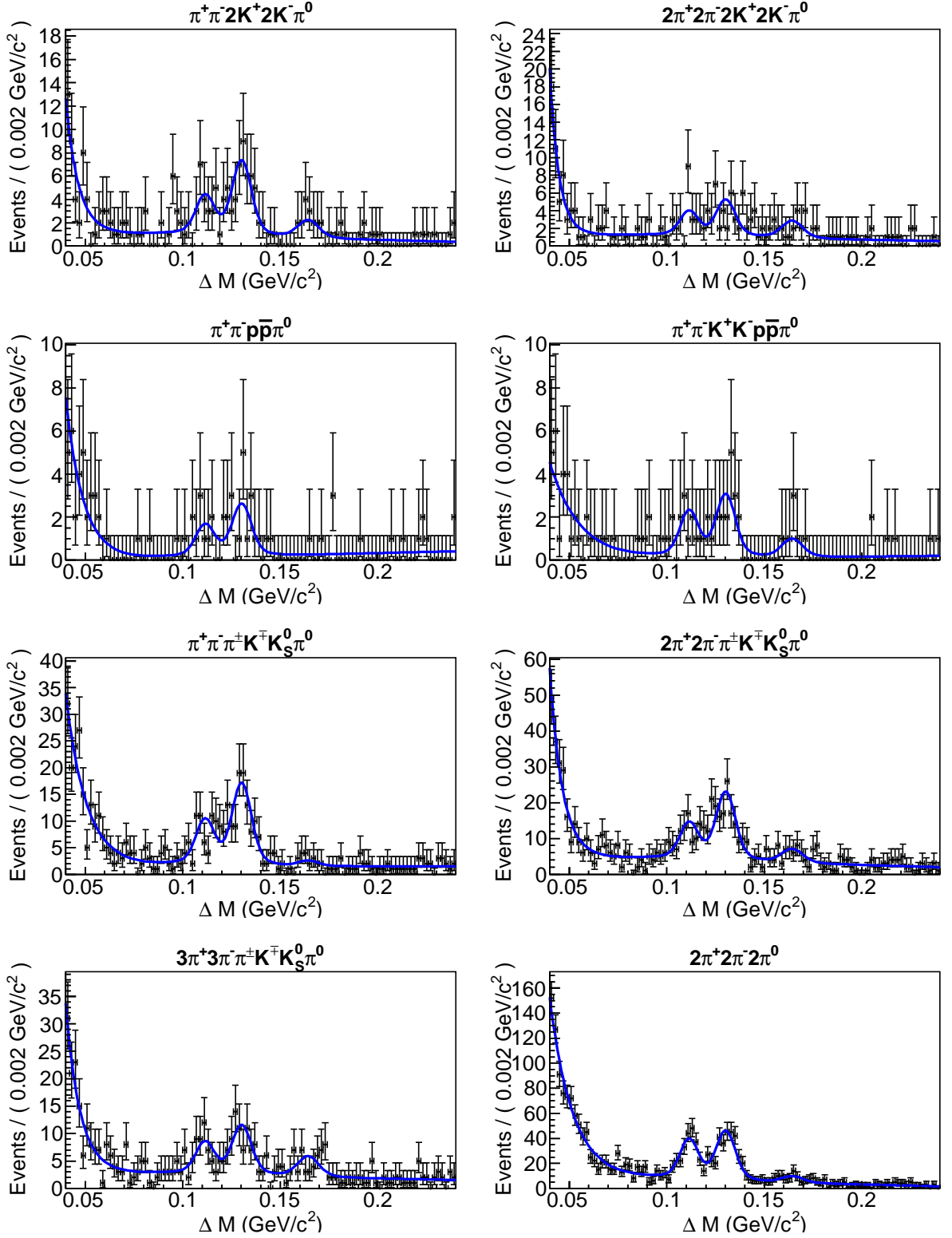}
  \caption{$\Delta M$ distributions in $\Y2S$ data for $\pi^+\pi^-2(K^+K^-)\pi^0$, $2(\pi^+\pi^-K^+K^-)\pi^0$, $\pi^+\pi^-p\antiproton\pi^0$, $\pi^+\pi^-K^+K^-p\antiproton\pi^0$, $\pi^+\pi^-\pi^{\pm}K^{\mp}\KS\pi^0$, $2(\pi^+\pi^-)\pi^{\pm}K^{\mp}\KS\pi^0$, $3(\pi^+\pi^-)\pi^{\pm}K^{\mp}\KS\pi^0$ and $2(\pi^+\pi^-)2\pi^0$ final states. Black dots with error bars are data and the blue curves represent the total fit result.}
  \label{fig:modes4}
\end{figure*}

\begin{figure*}[htp]
  \includegraphics[width=0.9\textwidth]{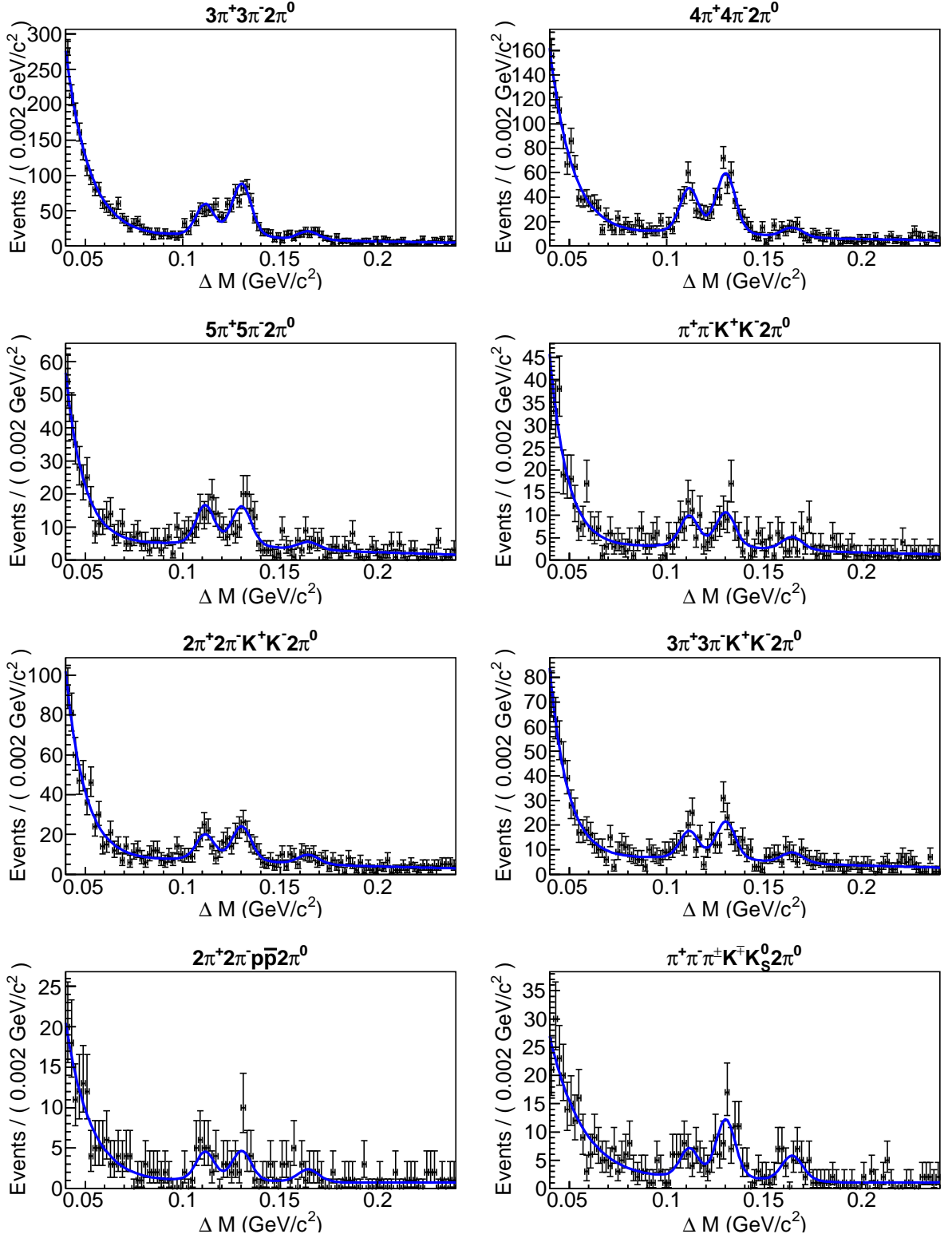}
  \caption{$\Delta M$ distributions in $\Y2S$ data for $3(\pi^+\pi^-)2\pi^0$, $4(\pi^+\pi^-)2\pi^0$, $5(\pi^+\pi^-)2\pi^0$, $(\pi^+\pi^-)K^+K^-2\pi^0$, $2(\pi^+\pi^-)K^+K^-2\pi^0$, $3(\pi^+\pi^-)K^+K^-2\pi^0$, $2(\pi^+\pi^-)p\antiproton 2\pi^0$ and $\pi^+\pi^-\pi^{\pm}K^{\mp}\KS 2\pi^0$ final states. Black dots with error bars are data and the blue curves represent the total fit result.}
  \label{fig:modes5}
\end{figure*}

\subsection{Branching Fraction Results}
\label{sec:mode:sys}
The major source of systematic uncertainty is due to the effect of possible 
intermediate decays (mentioned in Section~\ref{sec:data}) on the signal 
reconstruction efficiency. The deviation in efficiency relative to the default 
phase space assumption is 2 -- 23\%. Uncertainties on the signal yield due to 
PDF shapes are estimated by varying the shape parameters fixed in the nominal 
fit by $\pm 1\sigma$, and are found to be in the range 3 -- 12\%. The uncertainty due to the limited size of the signal MC sample is 1\%. Uncertainties associated with photon detection (3\%), charged track reconstruction (0.35\% per track), particle identification (1.5 -- 4.5\%), \KS reconstruction (2.2\% per \KS), $\pi^{0}$ reconstruction (2.2\% per $\pi^{0}$), and the number of \Y2S in data sample (2.3\%) are also taken into account. Systematic errors are added in quadrature mode by mode, and sum to between 6\% and 27\%.

The product branching fractions, 
${\cal B}[\Y2S\to\gamma\chi_{bJ}(1P)]\times{\cal B}[\chi_{b1}(1P)\to X_{i}]$, 
for each $\chi_{bJ}(1P)$ decay having significance exceeding $3\sigma$, are 
listed in Table~\ref{tab:BFstat} along with the corresponding statistical 
significance. In case of significance lower than $3\sigma$, we obtain an upper 
limit at $90\%$ confidence level (CL) on the branching fraction 
($\cal{B}_{\rm UL}$) by integrating the likelihood ($\cal{L}$)
of the fit with fixed values of the branching fraction: 
$\int_{0}^{\cal{B}_{\rm UL}} {\cal{L}}({\cal{B}}) d{\cal{B}}
=0.9\times \int_0^{1}{\cal{L}}({\cal{B}}) d{\cal{B}}$.
Systematic uncertainties in $\cal{B}_{\rm UL}$ are included by convolving the 
likelihood function with a Gaussian function with a width equal to the total 
uncertainty.
Our branching fraction results are consistent with, and more precise than, those reported in CLEO's analysis. Furthermore, a $\chi_{bJ}(1P)$ signal for 
$J=0,1$, and $2$ has been observed for the first time in 9, 27, and
16 modes, respectively. And, we also found first evidence of a $\chi_{bJ}(1P)$
 signal in 18, 1, and 14 modes for $J=0,1$, and $2$, respectively.
 \begin{table*}[htp]
   \caption{Product branching fractions 
     ${\cal B}[\Y2S\to\gamma\chi_{bJ}(1P)]\times{\cal B}[\chi_{b1}(1P)\to h_{i}]$
     ($\times 10^{-5}$) and statistical significance for each 
     $\chi_{bJ}(1P)(J=0,1,2)$ state. Upper limits at the 90\% CL are calculated
     for modes having significance less than 3$\sigma$. 
     The values indicated by the \d\ symbol represent the first observation or
     evidence of signal in that mode.}
  \label{sys1}
  \begin{tabular}
 {@{\hspace{0.25cm}}l@{\hspace{0.25cm}}  @{\hspace{0.25cm}}c@{\hspace{0.25cm}}
 @{\hspace{0.25cm}}l@{\hspace{0.25cm}}  @{\hspace{0.25cm}}c@{\hspace{0.25cm}}
 @{\hspace{0.25cm}}l@{\hspace{0.25cm}}  @{\hspace{0.25cm}}c@{\hspace{0.25cm}} @{\hspace{0.25cm}}c@{\hspace{0.25cm}}}
\hline \hline
\multirow{2}{*}{\bf Mode} & \multicolumn{2}{c}{{\boldmath$\chi_{b0}(1P)$}} & \multicolumn{2}{c}{{\boldmath$\chi_{b1}(1P)$}} & \multicolumn{2}{c}{{\boldmath$\chi_{b2}(1P)$}} \\
 & ${\cal B}$ & $\sigma$ & ${\cal B}$ & $\sigma$ & ${\cal B}$ & $\sigma$  \\
\hline
$2\pi^+2\pi^-$ & 0.13 $\pm$ 0.05 $\pm$ 0.02\d & 3.0 & 0.31 $\pm$ 0.06 $\pm$ 0.04\d & 6.8 & 0.15 $\pm$ 0.06 $\pm$ 0.02\d & 3.1 \\
$3\pi^+3\pi^-$ & 0.67 $\pm$ 0.08 $\pm$ 0.06\d & 11.0 & 1.84 $\pm$ 0.12 $\pm$ 0.16 & 23.6 & 0.96 $\pm$ 0.10 $\pm$ 0.10 & 12.6 \\
$4\pi^+4\pi^-$ & 0.78 $\pm$ 0.13 $\pm$ 0.11\d & 8.5 & 2.8 $\pm$ 0.2 $\pm$ 0.4 & 22.5 & 1.8 $\pm$ 0.2 $\pm$ 0.2 & 14.6 \\
$5\pi^+5\pi^-$ & 0.53 $\pm$ 0.14 $\pm$ 0.10\d & 4.9 & 1.5 $\pm$ 0.2 $\pm$ 0.3\d & 10.8 & 1.7 $\pm$ 0.2 $\pm$ 0.3\d & 10.9 \\
$\pi^+\pi^-K^+K^-$ & 0.15 $\pm$ 0.03 $\pm$ 0.03\d & 8.1 & 0.17 $\pm$ 0.03 $\pm$ 0.03\d & 8.6 & 0.15 $\pm$ 0.04 $\pm$ 0.03\d & 6.3 \\
$2\pi^+2\pi^-K^+K^-$ & 0.53 $\pm$ 0.08 $\pm$ 0.05 & 8.7 & 1.20 $\pm$ 0.11 $\pm$ 0.10 & 16.3 & 0.8 $\pm$ 0.10 $\pm$ 0.08 & 11.5 \\
$3\pi^+3\pi^-K^+K^-$ & 0.6 $\pm$ 0.13 $\pm$ 0.06 & 5.8 & 1.7 $\pm$ 0.2 $\pm$ 0.2 & 13.7 & 1.2 $\pm$ 0.2 $\pm$ 0.1\d & 9.8 \\
$4\pi^+4\pi^-K^+K^-$ & 1.2 $\pm$ 0.2 $\pm$ 0.2\d & 7.9 & 1.6 $\pm$ 0.2 $\pm$ 0.2\d & 10.5 & 1.6 $\pm$ 0.2 $\pm$ 0.2\d & 9.6 \\
$\pi^+\pi^-2K^+2K^-$ & 0.18 $\pm$ 0.05 $\pm$ 0.02\d & 5.4 & 0.35 $\pm$ 0.06 $\pm$ 0.03\d & 8.6 & 0.32 $\pm$ 0.07 $\pm$ 0.03\d & 7.4 \\
$2\pi^+2\pi^-2K^+2K^-$ & 0.33 $\pm$ 0.12 $\pm$ 0.03\d & 4.4 & 0.60 $\pm$ 0.12 $\pm$ 0.06\d & 7.8 & 0.56 $\pm$ 0.12 $\pm$ 0.06\d & 7.2 \\
$3\pi^+3\pi^-2K^+2K^-$ & 0.33 $\pm$ 0.12 $\pm$ 0.04\d & 3.8 & 0.42 $\pm$ 0.14 $\pm$ 0.06\d & 4.5 & 0.7 $\pm$ 0.2 $\pm$ 0.1\d & 6.2 \\
$2\pi^+2\pi^-p\overline{p}$ & $< 0.2 $ ~\ ~\ & 0.9 & 0.51 $\pm$ 0.08 $\pm$ 0.06\d & 10.2 & 0.16 $\pm$ 0.06 $\pm$ 0.03\d & 3.5 \\
$3\pi^+3\pi^-p\overline{p}$ & 0.23 $\pm$ 0.1 $\pm$ 0.03\d & 3.1 & 0.70 $\pm$ 0.14 $\pm$ 0.08\d & 7.8 & 0.31 $\pm$ 0.11 $\pm$ 0.04\d & 3.6 \\
$\pi^+\pi^-K^+K^-p\overline{p}$ & 0.13 $\pm$ 0.04 $\pm$ 0.02\d & 4.2 & 0.18 $\pm$ 0.05 $\pm$ 0.03\d & 5.7 & 0.15 $\pm$ 0.05 $\pm$ 0.03\d & 3.7 \\
$2\pi^+2\pi^-K^+K^-p\overline{p}$ & 0.31 $\pm$ 0.10 $\pm$ 0.05\d & 4.5 & 0.4 $\pm$ 0.1 $\pm$ 0.1\d & 6.3 & 0.2 $\pm$ 0.08 $\pm$ 0.03\d & 3.4 \\
$\pi^+\pi^-\pi^{\pm}K^{\mp}\KS$ & $< 0.1 $ ~\ ~\ & 0.0 & 0.7 $\pm$ 0.1 $\pm$ 0.1 & 12.9 & 0.28 $\pm$ 0.07 $\pm$ 0.05\d & 5.0 \\
$2\pi^+2\pi^-\pi^{\pm}K^{\mp}\KS$ & $< 0.4 $ ~\ ~\ & 2.2 & 1.9 $\pm$ 0.2 $\pm$ 0.2\d & 13.9 & 1.1 $\pm$ 0.2 $\pm$ 0.1\d & 8.5 \\
$3\pi^+3\pi^-\pi^{\pm}K^{\mp}\KS$ & $< 0.7 $ ~\ ~\ & 2.1 & 1.6 $\pm$ 0.3 $\pm$ 0.1\d & 8.9 & 0.8 $\pm$ 0.2 $\pm$ 0.1\d & 4.3 \\
$2\pi^+2\pi^-2\KS$ & 0.2 $\pm$ 0.08 $\pm$ 0.04\d & 4.2 & 0.28 $\pm$ 0.08 $\pm$ 0.03\d & 5.4 & 0.29 $\pm$ 0.09 $\pm$ 0.03\d & 5.2 \\
$3\pi^+3\pi^-2\KS$ & $< 0.6 $ ~\ ~\ & 2.2 & 0.5 $\pm$ 0.2 $\pm$ 0.1\d & 5.0 & 0.4 $\pm$ 0.2 $\pm$ 0.1\d & 3.8 \\
$\pi^+\pi^-K^+K^-\pi^0$ & $< 0.2 $ ~\ ~\ & 0.7 & 0.77 $\pm$ 0.10 $\pm$ 0.06 & 10.7 & 0.36 $\pm$ 0.09 $\pm$ 0.04 & 5.2 \\
$2\pi^+2\pi^-K^+K^-\pi^0$ & 0.8 $\pm$ 0.2 $\pm$ 0.2\d & 4.5 & 4.2 $\pm$ 0.3 $\pm$ 0.7 & 18.3 & 2.8 $\pm$ 0.3 $\pm$ 0.5 & 11.8 \\
$3\pi^+3\pi^-K^+K^-\pi^0$ & 1.8 $\pm$ 0.4 $\pm$ 0.4\d & 5.2 & 6 $\pm$ 0.6 $\pm$ 1.1 & 14.5 & 3.8 $\pm$ 0.5 $\pm$ 0.7 & 9.0 \\
$4\pi^+4\pi^-K^+K^-\pi^0$ & $< 1.7 $ ~\ ~\ & 1.4 & 4 $\pm$ 0.7 $\pm$ 1.0\d & 7.8 & 2.4 $\pm$ 0.6 $\pm$ 0.6\d & 4.4 \\
$\pi^+\pi^-2K^+2K^-\pi^0$ & 0.28 $\pm$ 0.11 $\pm$ 0.04\d & 3.2 & 0.9 $\pm$ 0.2 $\pm$ 0.2\d & 7.9 & 0.45 $\pm$ 0.15 $\pm$ 0.08\d & 3.9 \\
$2\pi^+2\pi^-2K^+2K^-\pi^0$ & 0.7 $\pm$ 0.3 $\pm$ 0.1\d & 3.2 & 1.1 $\pm$ 0.3 $\pm$ 0.2\d & 5.0 & 0.7 $\pm$ 0.3 $\pm$ 0.1\d & 3.5 \\
$\pi^+\pi^-p\overline{p}\pi^0$ & $< 0.1 $ ~\ ~\ & 0.0 & 0.24 $\pm$ 0.07 $\pm$ 0.04\d & 5.4 & 0.14 $\pm$ 0.06 $\pm$ 0.02\d & 3.3 \\
$\pi^+\pi^-K^+K^-p\overline{p}\pi^0$ & $< 0.5 $ ~\ ~\ & 2.8 & 0.5 $\pm$ 0.1 $\pm$ 0.2\d & 5.5 & 0.32 $\pm$ 0.13 $\pm$ 0.12\d & 3.3 \\
$\pi^+\pi^-\pi^{\pm}K^{\mp}\KS\pi^0$ & $< 0.5 $ ~\ ~\ & 1.5 & 2.2 $\pm$ 0.3 $\pm$ 0.2\d & 11.9 & 1.2 $\pm$ 0.2 $\pm$ 0.2\d & 6.1 \\
$2\pi^+2\pi^-\pi^{\pm}K^{\mp}\KS\pi^0$ & 1.3 $\pm$ 0.4 $\pm$ 0.2\d & 3.8 & 5.3 $\pm$ 0.6 $\pm$ 0.8 & 12.1 & 2.6 $\pm$ 0.5 $\pm$ 0.5\d & 6.1 \\
$3\pi^+3\pi^-\pi^{\pm}K^{\mp}\KS\pi^0$ & 2.4 $\pm$ 0.7 $\pm$ 0.5\d & 4.1 & 4.6 $\pm$ 0.8 $\pm$ 1.0\d & 7.6 & 2.9 $\pm$ 0.7 $\pm$ 0.6\d & 4.7 \\
$2\pi^+2\pi^-2\pi^0$ & 0.8 $\pm$ 0.2 $\pm$ 0.2\d & 3.9 & 4.5 $\pm$ 0.4 $\pm$ 1 & 16.9 & 3.4 $\pm$ 0.3 $\pm$ 0.8 & 12.5 \\
$3\pi^+3\pi^-2\pi^0$ & 3.6 $\pm$ 0.6 $\pm$ 0.5\d & 6.7 & 16.8 $\pm$ 0.9 $\pm$ 2.3 & 24.0 & 9.7 $\pm$ 0.9 $\pm$ 1.5 & 13.6 \\
$4\pi^+4\pi^-2\pi^0$ & 4.8 $\pm$ 1 $\pm$ 1.0\d & 5.3 & 22.3 $\pm$ 1.5 $\pm$ 4.7 & 19.6 & 15.5 $\pm$ 1.5 $\pm$ 3.3 & 13.3 \\
$5\pi^+5\pi^-2\pi^0$ & $< 5.1 $ ~\ ~\ & 2.6 & 10.8 $\pm$ 1.6 $\pm$ 2.4\d & 8.4 & 11 $\pm$ 1.9 $\pm$ 2.5\d & 7.1 \\
$\pi^+\pi^-K^+K^-2\pi^0$ & 0.5 $\pm$ 0.2 $\pm$ 0.1\d & 3.3 & 1.1 $\pm$ 0.2 $\pm$ 0.3\d & 7.0 & 0.9 $\pm$ 0.2 $\pm$ 0.2\d & 5.4 \\
$2\pi^+2\pi^-K^+K^-2\pi^0$ & 1.7 $\pm$ 0.5 $\pm$ 0.4\d & 3.9 & 4.9 $\pm$ 0.6 $\pm$ 1.1 & 10.0 & 3.5 $\pm$ 0.6 $\pm$ 0.8 & 6.8 \\
$3\pi^+3\pi^-K^+K^-2\pi^0$ & 3.2 $\pm$ 1 $\pm$ 0.8\d & 3.6 & 8.9 $\pm$ 1.2 $\pm$ 2.2\d & 9.4 & 6.4 $\pm$ 1.2 $\pm$ 1.6\d & 6.3 \\
$2\pi^+2\pi^-p\overline{p}2\pi^0$ & $< 1.8 $ ~\ ~\ & 2.7 & 1.8 $\pm$ 0.5 $\pm$ 0.3\d & 5.0 & 1.6 $\pm$ 0.5 $\pm$ 0.3\d & 4.4 \\
$\pi^+\pi^-\pi^{\pm}K^{\mp}\KS2\pi^0$ & 2.0 $\pm$ 0.5 $\pm$ 0.3\d & 5.1 & 3.6 $\pm$ 0.5 $\pm$ 0.4\d & 8.6 & 1.7 $\pm$ 0.5 $\pm$ 0.2 & 4.2 \\
$2\pi^+2\pi^-\pi^{\pm}K^{\mp}\KS2\pi^0$ & 3.0 $\pm$ 1.0 $\pm$ 0.6\d & 3.5 & 9 $\pm$ 1.3 $\pm$ 1.7\d & 9.1 & 5.1 $\pm$ 1.2 $\pm$ 1.0\d & 5.1 \\
\hline \hline
\end{tabular}
\label{tab:BFstat}
\end{table*}

\section{\boldmath $\chi_{b0}(1P)$ width measurement}
\label{sec:wid}
As mentioned in Section~\ref{sec:intro}, the $\chi_{b0}(1P)$ width may be above 
2~\mev, but no experimental measurement thus far has been attempted. 
The large signal yield obtained in our branching fraction studies of the 
$\chi_{bJ}(1P)$ triplet motivates a width measurement of the $\chi_{b0}(1P)$. 
For this study, we select the 41 significant modes described in 
Section~\ref{sec:mode:sys}. 
We obtain signal shape parameters from our high statistics MC sample, for which
the signal yield is 5 times the value observed for each mode in data. For 
fitting this MC sample each signal component is parameterized by a sum of a 
symmetric and an asymmetric Gaussian with common mean. The two widths of the 
asymmetric Gaussian and the fraction between the two Gaussians are identical 
for all the three signal components. 

\begin{figure}[htb]
  \includegraphics[width=0.5\textwidth]{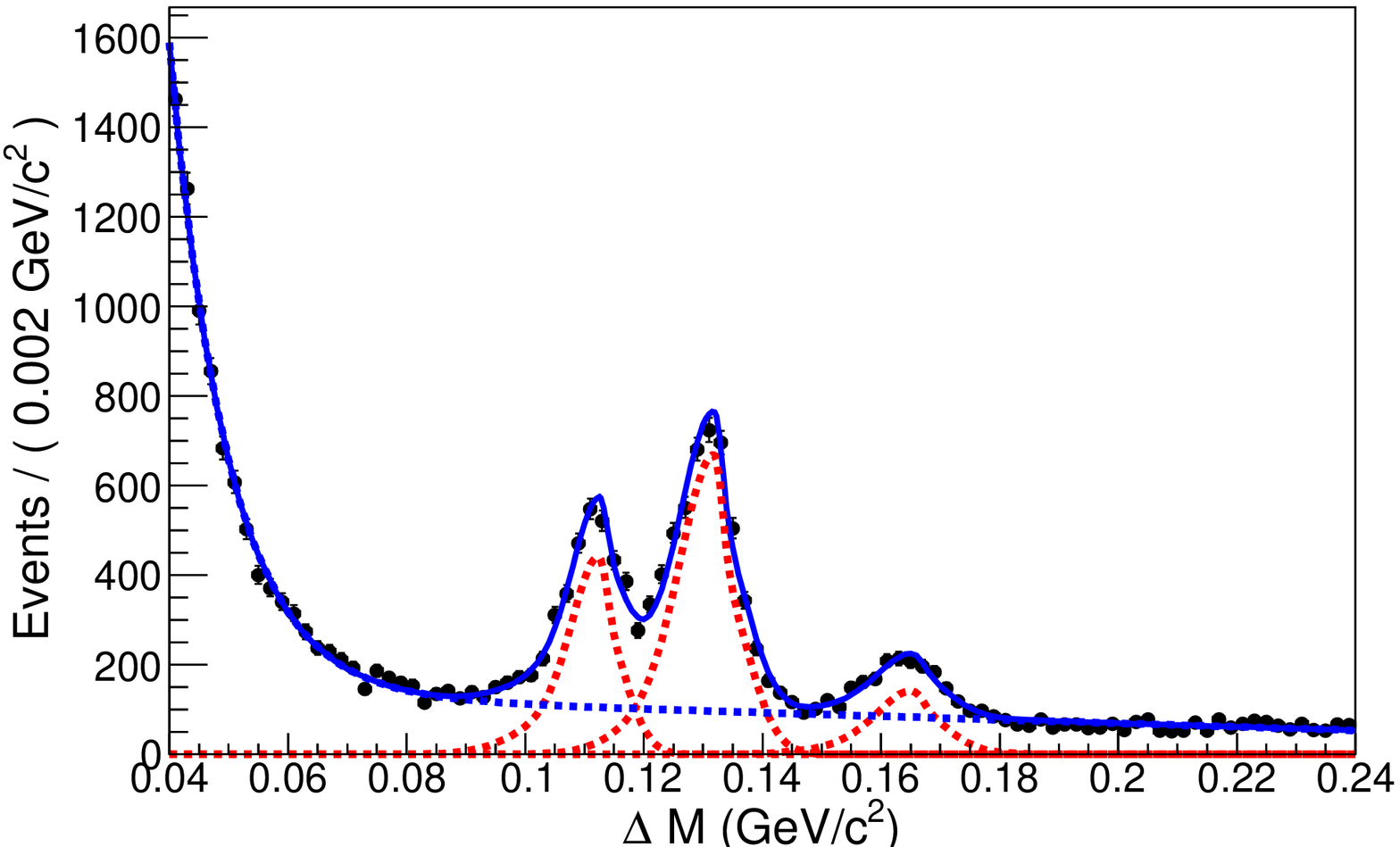}
  \caption{Result of maximum likelihood fit for width measurement to the 
    $\Delta M$ distribution in sum of 41 modes. 
    Black points with error bars are the data; blue solid and blue dashed curves
    are total fit and background components. 
    The three $\chi_{bJ}(1P)$ components are indicated by the red dashed curve.}
  \label{fig:wid_data}
\end{figure}

For the $\chi_{b0}(1P)$ width measurement in data, we model the three signals 
as described above for MC case, except for the $\chi_{b0}(1P)$, for which the 
sum of two Gaussian functions is convolved with a Breit-Wigner function. 
The symmetric part is convolved using the Voigtian function of 
RooFit~\cite{roofit}, whereas the asymmetric Gaussian function is convolved 
numerically using the FFTW (Fastest Fourier Transform in the West)~\cite{fftw} 
package of ROOT, with the same Breit-Wigner function whose width is floated. 
Individual widths of the symmetric Gaussians for all three signals are fixed to
the corresponding MC values, and are multiplied by a common 
resolution-correction factor to take account of possible data-MC difference. 
An unbinned extended maximum likelihood fit is performed in data to the 
$\Delta M$ distribution for the sum of the 41 modes having high significance. 
All background parameters components are floated, whereas all signal parameters
are fixed to the MC-fit values except for their individual mean and yield, and 
the resolution-correction factor. 
Figure~\ref{fig:wid_data} shows the result of the fit on data. 
The $\chi_{b0}(1P)$ width is found to be 1.3 $\pm$ 0.9\,\mev. 
In the absence of a statistically significant result, we derive an upper limit 
on the width (see below). 

The systematic uncertainity due to the fixed PDF parameters is $\pm0.2\mev$. 
To estimate uncertainty due to the assumption of negligible $\chi_{b2}(1P)$ 
width, a non-zero width (0.5~\mev) to $\chi_{b2}(1P)$  is introduced and this 
affects $\chi_{b0}(1P)$ width by $\pm0.1\mev$.
We generate a large ensemble of pseudo-experiments for different $\chi_{b0}(1P)$
width hypotheses using the PDF parameters obtained from data, and perform a 
linearity test between the generated and fitted width values. An uncertainty of
$\pm0.1\mev$ is assigned to account for possible deviation from linearity.
We obtain a $90\%$ CL upper limit on the width of the $\chi_{b0}(1P)$ 
($\cal{W}_{\rm limit}$) by integrating the likelihood ($\cal{L}$) of the fit using
fixed values of the width:  
$\int_{0}^{\cal{W}_{\rm limit}} {\cal{L}} ({\cal{W}}) d {\cal{W}} = 0.9 \times \int_{0}^{\infty} {\cal{L}} ({\cal{W}}) d{\cal{W}}$. 
Systematic uncertainties are included by convolving the likelihood function with
a Gaussian function of width equal to the total systematic uncertainty. 
We estimate the width of the $\chi_{b0}(1P)$ $< 2.4 \mev$. 

\section{Summary}
We have studied $\chi_{bJ}(1P)$ states produced from the $\Y2S$ via electric 
dipole radiative transition, decaying to light hadronic final states. 
Our measurements are consistent with, and more precise than, those reported by 
the CLEO Collaboration~\cite{chibj1p:cleo}. 
We also report a $\chi_{bJ}(1P)$ signal for $J=0,1$, and $2$ for the first time
in 27, 28, and 30 modes, respectively. 
Furthermore, in the absence of a statistically significant result, a $90\%$ confidence-level upper limit is set on the $\chi_{b0}(1P)$ width at $\Gamma_{\rm total}< 2.4\mev$. 

\section*{Acknowledgments}
We thank the KEKB group for the excellent operation of the
accelerator; the KEK cryogenics group for the efficient
operation of the solenoid; and the KEK computer group,
the National Institute of Informatics, and the 
PNNL/EMSL computing group for valuable computing
and SINET4 network support.  We acknowledge support from
the Ministry of Education, Culture, Sports, Science, and
Technology (MEXT) of Japan, the Japan Society for the 
Promotion of Science (JSPS), and the Tau-Lepton Physics 
Research Center of Nagoya University; 
the Australian Research Council;
Austrian Science Fund under Grant No.~P 22742-N16 and P 26794-N20;
the National Natural Science Foundation of China under Contracts 
No.~10575109, No.~10775142, No.~10875115, No.~11175187, No.~11475187
and No.~11575017;
the Chinese Academy of Science Center for Excellence in Particle Physics; 
the Ministry of Education, Youth and Sports of the Czech
Republic under Contract No.~LG14034;
the Carl Zeiss Foundation, the Deutsche Forschungsgemeinschaft, the
Excellence Cluster Universe, and the VolkswagenStiftung;
the Department of Science and Technology of India; 
the Istituto Nazionale di Fisica Nucleare of Italy; 
the WCU program of the Ministry of Education, National Research Foundation (NRF) 
of Korea Grants No.~2011-0029457,  No.~2012-0008143,  
No.~2012R1A1A2008330, No.~2013R1A1A3007772, No.~2014R1A2A2A01005286, 
No.~2014R1A2A2A01002734, No.~2015R1A2A2A01003280 , No. 2015H1A2A1033649;
the Basic Research Lab program under NRF Grant No.~KRF-2011-0020333,
Center for Korean J-PARC Users, No.~NRF-2013K1A3A7A06056592; 
the Brain Korea 21-Plus program and Radiation Science Research Institute;
the Polish Ministry of Science and Higher Education and 
the National Science Center;
the Ministry of Education and Science of the Russian Federation and
the Russian Foundation for Basic Research;
the Slovenian Research Agency;
Ikerbasque, Basque Foundation for Science and
the Euskal Herriko Unibertsitatea (UPV/EHU) under program UFI 11/55 (Spain);
the Swiss National Science Foundation; 
the Ministry of Education and the Ministry of Science and Technology of Taiwan;
and the U.S.\ Department of Energy and the National Science Foundation.
This work is supported by a Grant-in-Aid from MEXT for 
Science Research in a Priority Area (``New Development of 
Flavor Physics'') and from JSPS for Creative Scientific 
Research (``Evolution of Tau-lepton Physics'').

\end{document}